\documentclass[amsmath,amssymb,amsbsy,prb,twocolumn,preprintnumbers,showpacs]{revtex4-2}
\usepackage{color}
\usepackage{dcolumn}
\usepackage{bm}
\usepackage{braket}
\usepackage{mathtools}
\usepackage{ulem}
\usepackage{mathrsfs}
\usepackage{times}
\usepackage[breaklinks,colorlinks=true,linkcolor=blue,urlcolor=blue,citecolor=blue]{hyperref}

\begin{document}
\title{Square-root topological semimetals}
\author{Tomonari Mizoguchi}
\affiliation{Department of Physics, University of Tsukuba, Tsukuba, Ibaraki 305-8571, Japan}
\email{mizoguchi@rhodia.ph.tsukuba.ac.jp}
\author{Tsuneya Yoshida}
\affiliation{Department of Physics, University of Tsukuba, Tsukuba, Ibaraki 305-8571, Japan}
\author{Yasuhiro Hatsugai}
\affiliation{Department of Physics, University of Tsukuba, Tsukuba, Ibaraki 305-8571, Japan}

\begin{abstract}
We propose topological semimetals generated by the square-root operation for tight-binding models in two and three dimensions, 
which we call square-root topological semimetals.
The square-root topological semimetals host topological band touching at finite energies, 
whose topological protection is inherited from the squared Hamiltonian. 
Such a topological character is also reflected in emergence of boundary modes with finite energies.
Specifically, focusing on topological properties of squared Hamiltonian in class AIII,  we reveal that a decorated honeycomb (decorated diamond) model hosts finite-energy Dirac cones (nodal lines).
We also propose a realization of a square-root topological semimetal in a spring-mass model, where 
robustness of finite-energy Dirac points against the change of tension is elucidated. 
\end{abstract}
\maketitle

\section{Introduction}  
In the past decade, novel classes of topological phases have been extensively explored~\cite{Wen2017}. 
Focusing on non-interacting fermions, there are two kinds of topological phases according to the bulk spectrum.
One is a gapped topological phase where bulk has an energy gap
and nontrivial topological numbers are defined for Bloch or Bogoliubov bands.
Examples include
topological insulators (TIs)~\cite{Haldane1988,Kane2005,Kane2005_2,Bernevig2006,Hasan2010,Qi2011} 
and topological superconductors (TSCs)~\cite{Kitaev2001,Fu2008,Fu2010,Mizushima2016,Sato2016,Sato2017}. 
In TIs and TSCs, the nontrivial topology is known to result in robust boundary modes;
this relation is called bulk-boundary correspondence~\cite{Hatsugai1993,Hatsugai1993_2}.  
The other kind of topological phase is a gapless topological phase, also termed
a topological semimetal (TSM),~\cite{Murakami2007,Burkov2011,Vafek2014,Burkov2016,Armitage2018}
where bulk bands themselves have gapless points or nodes, protected by nontrivial topology. 
Topologically-protected boundary modes appear in TSMs as well. 
For instance, flat edge modes
protected by the winding number
appear in graphene nano-ribbon with the zigzag edge~\cite{Fujita1996,Ryu2002,Hatsugai2009} 
and the $d_{x^2-y^2}$ superconductor~\cite{Hu1994,Tanaka1995}.

Recently, an interesting proposal to obtain a class of TIs was made by Arkinstall, et al.
Their proposal is to take the square-root of topological tight-binding Hamiltonians~\cite{Arkinstall2017}.
TIs thus obtained are called square-root TIs. 
The square-root operation was introduced in several contexts,
such as a correspondence between the bosonic Klien-Gordon theory and the fermionic Dirac theory~\cite{Dirac1928}.
Recently, the similar argument was also applied to the spin models~\cite{Attig2017} and mechanical systems~\cite{Kane2014,Attig2019}.
Concerning the tight-binding models, the square-root operation
is carried out by adding the ``mediating sites" between sites of original lattice having non-zero particle transfer, 
and letting the nearest-neighbor (NN) transfer between the original sites and mediating sites.
This implies that the model is naturally chiral symmetric 
as the bipartition of the entire lattice to the original sites and the mediating sites is possible. 
The square of the model is equal to the direct sum of the original model and another model defined on mediating sites (up to the constant shift). 
When topological models are set as an original Hamiltonian, its square-root inherits the topological nature of it.
So far, tight-binding models of TIs~\cite{Arkinstall2017,Kremer2020}, higher-order TIs~\cite{Mizoguchi2020_sq}, and  non-Hermitian TIs~\cite{Ezawa2020} created by the square-root operation were proposed.
Their experimental realization was also actively pursued
in various artificial materials, such as
photonic crystals~\cite{Arkinstall2017,Kremer2020}, 
electric circuits~\cite{Song2020}, and phononic crystals~\cite{Yan2020}.
Interestingly, in square-root TIs preserving the chiral symmetry, topologically-protected boundary modes appear at positive and negative energies in a pairwise manner, which reflects the square-root nature of the model.
Along with the above significant progress in insulators and superconductors, to our best knowledge, square-root topology of semimetals has not been explored yet. 

In this paper,
we propose that TSMs can also be generated by the square-root operation.
We term such TSMs square-root TSMs (SR-TSMs).
SR-TSMs have topological band touching at finite energies 
and the topologically-protected nature of them can be elucidated by considering a squared Hamiltonian. 

As concrete examples, we study a series of $d$-dimensional decorated diamond model,
i.e., a decorated honeycomb model in two dimensions and a decorated diamond model in three dimensions.
Their mediating sites correspond to the vertices of line graphs~\cite{Mielke1991,Mielke1991_2}, 
i.e., the $d$-dimensional pyrochlore lattice~\cite{Hatsugai2011}.
The models are a suitable platform for the SR-TSMs 
since the squared Hamiltonian is composed of $d$-dimensional diamond and 
pyrochlore lattices, both of which are well-known examples of the TSM. (The former is in class AIII.)
Specifically, the Dirac points arise in two dimensions (i.e., the hoeycomb and kagome lattices), 
and the nodal line SM is realized in three dimensions (i.e., the diamond and pyrochlre lattices). 
We show that the $d$-dimensional decorated diamond model indeed hosts the nodal points or lines at finite energies,
inherited from the $d$-dimensional diamond and pyrochlore models.
Furthermore, the decorated models also succeed to
the characteristic boundary modes in the $d$-dimensional diamond model,
i.e., the flat edge modes in the two-dimensional case and the flat surface state in the three-dimensional case. 
They appear at finite energies, and
are topologically protected by the winding number defined for the $d$-dimensional diamond-lattice of the sector squared Hamiltonian. 

We further propose that the SR-TSM is feasible in a spring-mass model~\cite{Prodan2009,Kane2014,Kariyado2015,Susstrunk2015,Huber2016,Kariyado2016,Po2016,Socolar2017,Takahashi2017,SerraGarcia2018,Takahashi2019,Yoshida2019,Attig2019,Wakao2020,Wakao2020_2}, a periodic array of mass points connected by springs.
Focusing on the two-dimensional case, i.e., the decorated honeycomb spring-mass model,
we reveal that the finite-energy band touching points 
appear and they survive when the inter-mode coupling between longitudinal and transverse modes is introduced.
This indicates that those touching points are topologically stable as far as the protecting symmetry is conserved.

The rest of this paper is structured as follows.
We first argue a generic recipe for constructing SR-TSMs of class AIII in Sec.~\ref{sec:gen}.
Then, in the following two sections, we reveal how this construction works through the concrete examples.
In Sec.~\ref{sec:DH}, we study the decorated honeycomb model as an example of 
SR Dirac semimetals in two dimensions.
In Sec.~\ref{sec:DD}, we study the decorated diamond model, as an example of SR nodal line semimetals in three dimensions.
Section~\ref{sec:SMmodel} is devoted to the spring-mass-model realization of the SR-TSM. 
In Sec.~\ref{sec:summary}, we summarize this paper.

\section{Generic construction of square-root topological semimetals\label{sec:gen}}
In this section, we describe a generic recipe of constructing SR-TSMs, focusing on those in class AIII.

Consider a bipartite lattice with an even number of sublattices and 
vertices of its line graph, which are obtained by
placing a site on each bond of the original lattice.
Then, let us consider the Hamiltonian with the NN hoppings
on a composite lattice of the original lattice and the vertices of its line graph:
\begin{eqnarray}
H = \sum_{\bm{k}} \bm{c}^\dagger_{\bm{k}} \mathscr{H}_{\bm{k}} \bm{c}_{\bm{k}}, 
\end{eqnarray}
where $\bm{c}_{\bm{k}} = \left(c_{\bm{k},1}, \cdots c_{\bm{k},N}, c_{\bm{k},N+1},\cdots,c_{\bm{k},N+M} \right)^{\rm T}$.
Here, $N$ and $M$ are the numbers of the sublattices for the original lattice and 
the vertices of the line graph, respectively, and 
 the sublattices of the composite lattice are labeled such that
the sublattices 1-$N$ belong to the original lattice 
whereas the sublattice $(N+1)$-$(N+M)$ to the vertices of the line graph.

This kind of lattice structure is a suitable platform for realizing square-root topological phases.
For instance, the square-root higher-order TI was proposed~\cite{Mizoguchi2020_sq},
as was a generic construction of square-root TIs proposed later on~\cite{Ezawa2020}.
As the NN hoppings on the composite lattice occurs only between the 
original lattice and the vertices of the line graph, 
the Hamiltonian matrix $\mathscr{H}_{\bm{k}}$ can be written in a form: 
\begin{eqnarray}
\mathscr{H}_{\bm{k}}  = 
\begin{pmatrix}
\mathcal{O}_{N,N} &t  \Psi_{\bm{k}}^\dagger \\
t  \Psi_{\bm{k}} &\mathcal{O}_{M,M}  \\
\end{pmatrix}. \label{eq:Ham_gen}
\end{eqnarray}
Here, $t$ is the transfer integral, 
$ \Psi_{\bm{k}} $ is the $M \times N$ matrix which reflects the connectivity between 
the original lattice and the vertices of the line graph,  
and $\mathcal{O}_{n,m}$ stands for the $n\times m$ zero matrix. 

From (\ref{eq:Ham_gen}), we see that the model preserves the chiral symmetry, 
namely, $\mathscr{H}_{\bm{k}}$ anti-commutes a matrix $\Gamma$:
\begin{eqnarray}
\Gamma = \begin{pmatrix}
I_{N} & \mathcal{O}_{N,M}\\
\mathcal{O}_{M,N} & - I_M \\
\end{pmatrix}, \label{eq:Gamma}
\end{eqnarray}
where $I_ n$ is the $n\times n$ identity matrix. 
Due to this chiral symmetry, 
the square of $\mathscr{H}_{\bm{k}} $ is block-diagonalized,
since $\left(\mathscr{H}_{\bm{k}}\right)^2$ commutes $\Gamma$.
Specifically, we have
\begin{eqnarray}
\left(\mathscr{H}_{\bm{k}} \right)^2  = 
\begin{pmatrix}
t^2 \Psi_{\bm{k}}^\dagger\Psi_{\bm{k}} &\mathcal{O}_{N,M}  \\
\mathcal{O}_{M,N} &t^2  \Psi_{\bm{k}}\Psi_{\bm{k}}^\dagger \\
\end{pmatrix}.
\end{eqnarray}
Notably, $t^2 \Psi_{\bm{k}}^\dagger\Psi_{\bm{k}}$ corresponds to 
the Hamiltonian on the original lattice with the on-site potential $z t^2$ 
and the NN hopping $t^2$,
whereas $t^2 \Psi_{\bm{k}} \Psi_{\bm{k}}^\dagger$ corresponds to that on the line graph with the on-site potential $z^\prime t^2$ 
and the NN hopping $t^2$~\cite{Hatsugai2011}; $z$ and $z^\prime$ are the coordination numbers of the original sites and the vertices of the line graph, respectively.
The eigenenergies of these two matrices are identical except for the zero-energy flat band in that for the line graph. 
In addition, we can define the topological winding number for the squared Hamiltonian in the original lattice subspace. 
To be specific, as $N$ is even, there exists a $N \times N$ matrix $\Sigma$ 
such that $\Sigma$ satisfies
\begin{eqnarray}
\{ \Sigma, \tilde{h}_{\bm{k}} \} = 0, \label{eq:sigma}
\end{eqnarray}
with 
\begin{eqnarray}
\tilde{h}_{\bm{k}} :=t^2 \Psi_{\bm{k}}^\dagger\Psi_{\bm{k}}-zt^2 I_{N},
\end{eqnarray}
and $\{, \}$ stands for the anitcommutator.
Then, the following winding number can be defined~\cite{Ryu2002,Hatsugai2009}: 
\begin{eqnarray}
\nu (k_1,\cdots, k_{j-1},k_{j+1}, \cdots)= \int_0^{1} \frac{d k_j}{4\pi i} \mathrm{Tr} 
\left[\Sigma \partial_{k_j} \left( \log \tilde{h}_{\bm{k}} \right) \right], \nonumber \\ \label{eq:winding_gen}
\end{eqnarray}
where $k_j \in [0,1]$ is defined such that $\bm{k} = \sum_{j=1}^d k_j \bm{b}_j$ 
with $d$ being the spatial dimension and $\bm{b}_j$ being the $j$-th reciprocal lattice vector.

The dispersion relation of the Hamiltonian of the composite lattice is given by the 
square of that for $\left(\mathscr{H}_{\bm{k}} \right)^2$, with the signs $+$ and $-$. 
This means that, if the NN hopping model on the original lattice has band touching at zero energy,
the Hamiltonian $\mathscr{H}_{\bm{k}}$ has the band touchings as well, and their energies 
are $\pm \sqrt{z} |t|$, i.e., the TSM is realized in the composite-lattice model.
However, we cannot define the topological winding number protecting 
the band touchings from the composite-lattice model itself, as they are at finite energies. 
Instead, they are protected by the winding number for the squared Hamiltonian defined in Eq.~(\ref{eq:winding_gen}). 
Accordingly, the protecting symmetry of the finite-energy gapless nodes is the chiral symmetry described by $\Sigma$ of Eq.~(\ref{eq:sigma})
because this symmetry enables us to define the winding number.
We thus call this TSM the SR-TSM. 
In addition, the topologically protected band touching also leads to the finite-energy boundary modes,
which is another characteristics of the SR-TSM.

In the following two sections, we show the concrete models,
namely, the decorated honeycomb model in two dimensions and the decorated diamond model in three dimensions. 
Additionally, we remark that the SR-TSM obtained in the above procedures is stable against the on-site potential 
that is proportional to $\Gamma$; see Appendix~\ref{sec:onsite} for details. 

\section{Example 1: square-root Dirac semimetal in the decorated honeycomb model \label{sec:DH}}
\begin{figure}[b]
\begin{center}
\includegraphics[clip,width = 0.98\linewidth]{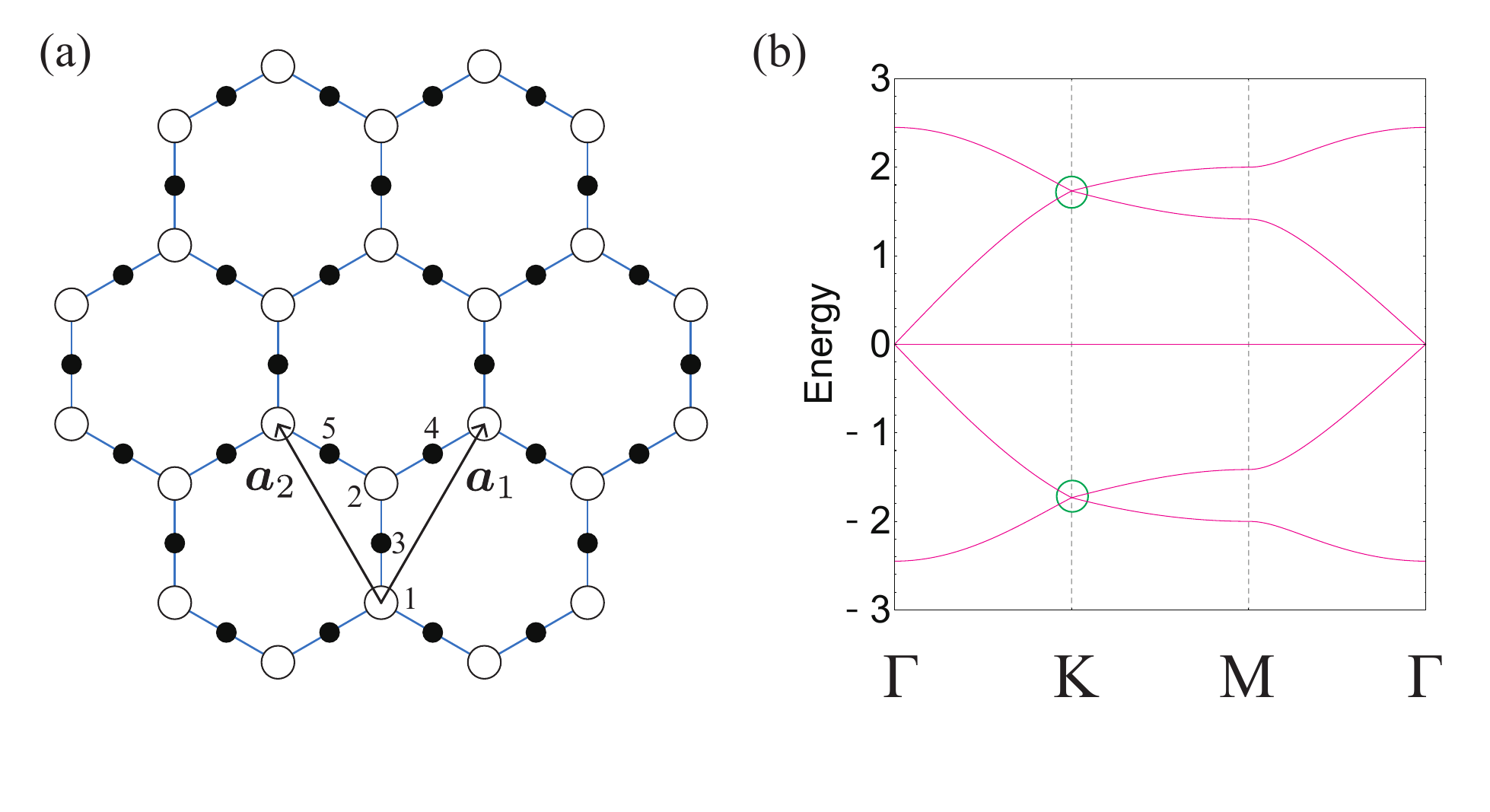}
\caption{(a) A decorated honeycomb lattice and (b) the band structure with $t = 1$.
The finite-energy Dirac points are denoted by green circles.}
  \label{fig:model_dh}
 \end{center}
\end{figure}
\subsection{Model and bulk properties}
We first study the decorated honeycomb model,
which is a composite lattice of honeycomb and kagome lattices [Fig.~\ref{fig:model_dh}(a)].
This model was studied in the literature~\cite{Shima1993,Barreteau2017,Lee2020}, 
and the aim of this paper is to present a renewed viewpoint of the SR-TSM.

The lattice has five sublattices degrees of freedoms; 
two of them come from the honeycomb lattice and 
the rest from the kagome lattice. 
Then, $\Psi_{\bm{k}}$ of Eq.~(\ref{eq:Ham_gen}) is a $3 \times 2$ matrix given as
\begin{eqnarray}
\Psi_{\bm{k}} = 
\begin{pmatrix}
1 & 1 \\
e^{i\bm{k}\cdot \bm{a}_1} & 1 \\
e^{i\bm{k}\cdot \bm{a}_2} & 1 \\
\end{pmatrix}.  \label{eq:psih}
\end{eqnarray}
Note that the two lattice vectors in Eq.~(\ref{eq:psih})
are $\bm{a}_1 = \left( \frac{1}{2},  \frac{\sqrt{3}}{2}\right)$ and $\bm{a}_2 = \left(-\frac{1}{2},  \frac{\sqrt{3}}{2}\right)$; the corresponding reciprocal lattice vectors are 
$\bm{b}_1 = \left(2\pi,  \frac{2\pi}{\sqrt{3}}\right)$ and $\bm{b}_2 = \left(- 2\pi,  \frac{2\pi}{\sqrt{3}}\right)$. 

As explained in Sec.~\ref{sec:gen}, the square of $\mathscr{H}_{\bm{k}}$ 
is block-diagonalized as 
\begin{eqnarray}
\mathscr{H}_{\bm{k}}^2  = 
\begin{pmatrix}
\mathscr{H}_{\bm{k}}^{(\mathrm{H})} &\mathcal{O}_{2,3}  \\
\mathcal{O}_{3,2} & \mathscr{H}_{\bm{k}}^{(\mathrm{K})}  \\
\end{pmatrix},
\end{eqnarray}
where $\mathscr{H}_{\bm{k}}^{(\mathrm{H})}$ and $\mathscr{H}_{\bm{k}}^{(\mathrm{K})}$
correspond to the honeycomb lattice model with the NN hopping $t^2$ and the on-site potential $3t^2$
and the kagome lattice model with the NN hopping $t^2$ and the on-site potential $2t^2$,
respectively.
For later use, we define
\begin{eqnarray}
q^{\rm (H)}_{\bm{k}} = 1 + e^{ i\bm{k}\cdot \bm{a}_1} + e^{i\bm{k}\cdot \bm{a}_2},
\end{eqnarray}
which is the $(2,1)$ component of $\mathscr{H}_{\bm{k}}^{(\mathrm{H})}$. 

\begin{figure}[t]
\begin{center}
\includegraphics[clip,width = 0.95\linewidth]{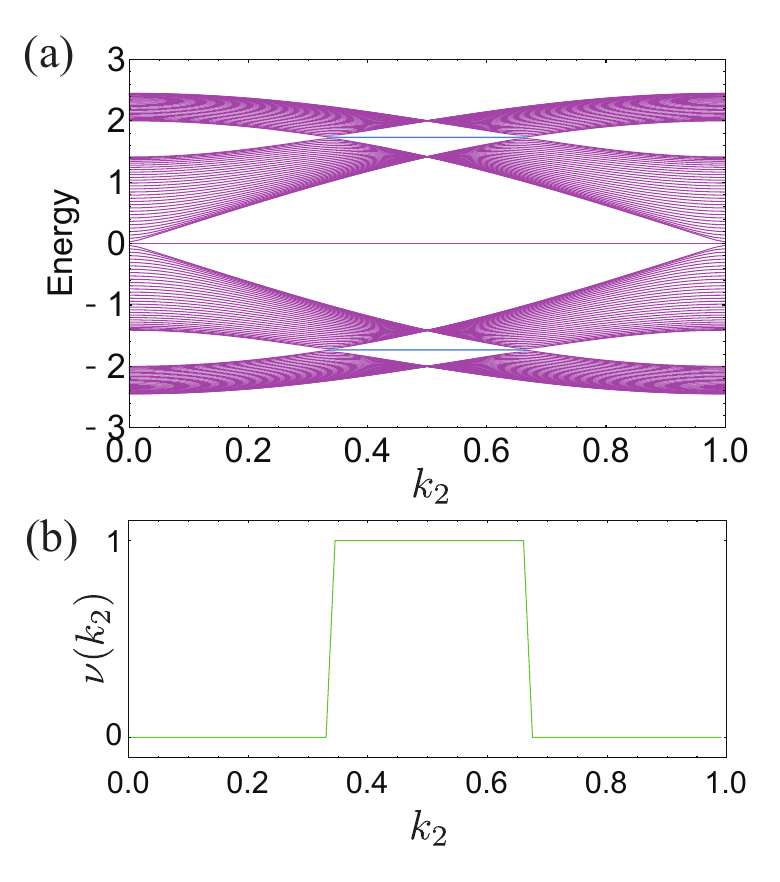}
\caption{(a) The dispersions for the system with the zigzag edge.
The cyan lines represent the flat edge modes. 
(b) The winding number defined for $q^{\rm (H)}_{\bm{k}}$ in Eq.~(\ref{eq:winding_H}).
We set $t = 1$.}
  \label{fig:edge}
 \end{center}
\end{figure}

The band structure for this model is shown in Fig~\ref{fig:model_dh}(b).
We find Dirac cones at K and K$^\prime$ points (the latter is not shown in the figure), whose 
energies are $\varepsilon = \pm \sqrt{3} |t|$.
These Dirac cones are inherited from those of the honeycomb and kagome models.
We note that the band gap opens when 
the hopping between sublattice 1 and 3,4,5 is different from that between sublattice 2 and 3,4,5,
because the chiral symmetry of the squared Hamiltonian [Eq.~(\ref{eq:sigma})] is broken.
In such a case, higher-order TI is realized~\cite{Mizoguchi2020_sq}.

\subsection{Edge modes and topological protection}
Figure~\ref{fig:edge}(a) shows the dispersion relation for the cylinder,
obtained by assigning the open boundary condition in the direction of $\bm{a}_1$.
The edge shape is chosen to be the zigzag edge.
In the momentum space, $k_2$ defined in Sec.~\ref{sec:gen} remains as a good quantum number. 
We see flat edge modes at $\varepsilon = \pm \sqrt{3} |t|$, 
connecting the Dirac points in the bulk;
this is reminiscent of the edge modes of the conventional honeycomb model under the zigzag edge,
besides the fact that their energies are finite rather than zero. 

The finite-energy flat edge modes are topologically protected 
by the winding number for the honeycomb sector of the squared Hamiltonian, as we discussed in Sec.~\ref{sec:gen}.
Specifically,
we can define the following winding number as a function of $k_2$: 
\begin{eqnarray}
\nu (k_2) =\frac{1}{2\pi i} \int_0^1 d k_1 \hspace{0.5mm} \frac{\partial_{k_1} q^{\rm (H)}_{\bm{k}}}{q^{\rm (H)}_{\bm{k}}}. \label{eq:winding_H}
\end{eqnarray}
Note that the definition of Eq.~(\ref{eq:winding_H}) coincides with that of Eq.~(\ref{eq:winding_gen}) 
when setting $\Sigma = \mathrm{diag} (1,-1)$.
This winding number is exactly identical to that for the honeycomb model, but we perform the integration in Eq.~(\ref{eq:winding_H}) for completeness.
Changing the variable as $z = e^{2 \pi i k_1}$, we have
\begin{eqnarray}
\nu (k_2) =\frac{1}{2\pi i} \oint_C d z \hspace{0.5mm} \frac{1} {z + 1 + e^{2\pi i k_2}}, \label{eq:winding_H_2}
\end{eqnarray}
where $C$ is a unit circle in the complex plane.
Using Cauchy's residue theorem, we find that  
$\nu (k_2)$ is 0 (1) if $|1 + e^{2\pi i k_2}| > 1$ ( $|1 + e^{2\pi i k_2}| < 1$). 
At the critical point, where $|1 + e^{2\pi i k_2}| = 1$, (i.e., $k_2 = \frac{1}{3}, \frac{2}{3}$) 
the band gap closes at certain $k_1$, which is nothing but K and K$^\prime$ points~\cite{Remark}.
In fact, this also manifests the topological stability of the Dirac points in a bulk,
because the winding number defined by the contour integral around the Dirac point takes non-trivial value~\cite{Wen1989,Hatsugai2009,Ryu2002}. 
The bulk-edge correspondence tells us that the number of 
zero-energy boundary mode for $\mathscr{H}_{\bm{k}}^{(\mathrm{H})} -3t^2$ (per edge) is equal to $\nu (k_2)$ at each $k_2$.
Then, by taking into account the constant shift of $3t^2$, we find that 
the edge modes with energies $\pm \sqrt{3} |t|$ arise in the decorated honeycomb model.
In Fig.~\ref{fig:edge}(b), we plot $\nu (k_2)$, which clearly coincides with the above argument, 
in that $\nu (k_2)$ is one where the edge modes exist, while it is zero otherwise. 
From this result, we conclude that the SR Dirac semimetal is realized in the decorated honeycomb model.

Here we note yet another way of characterizing the bulk Dirac points and the edge modes, 
i.e., using the Berry's phase~\cite{Delplace2011,Kariyado2013}.
In fact, such a characterization is applied to the honeycomb model~\cite{Delplace2011},
namely, the Berry's phase is quantized in $\mathbb{Z}_2$ and it takes $\pi$ ($0$) where the edge modes do (or do not) exist. 
Considering the fact that the squared Hamiltonian is identical to that of the honeycomb model,
we find in the decorated honeycomb model that the Berry's phase 
defined for the subspace of the honeycomb lattices is indeed quantized in $\mathbb{Z}_2$,
and it also gives the topological characterization of the edge modes. 

\section{Example 2: Square-root nodal-line semimetal in the decorated diamond model \label{sec:DD}}
\begin{figure}[t]
\begin{center}
\includegraphics[clip,width = 0.98\linewidth]{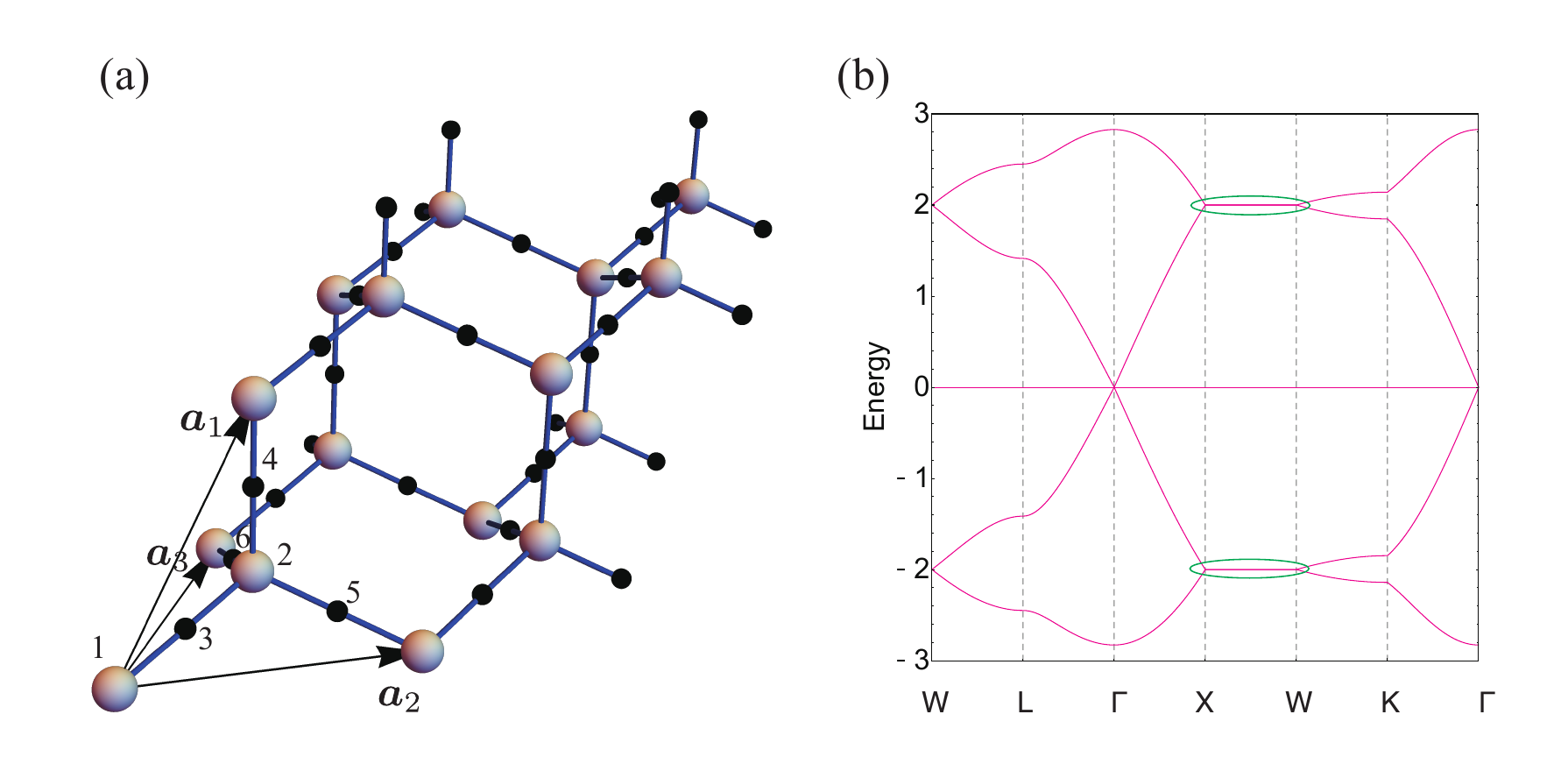}
\caption{(a) A decorated diamond lattice and (b) the band structure with $t=1$. 
The finite-energy nodal lines are denoted by green ellipses.}
  \label{fig:sd_bulk}
 \end{center}
\end{figure}
\begin{figure}[b]
\begin{center}
\includegraphics[clip,width = 0.98\linewidth]{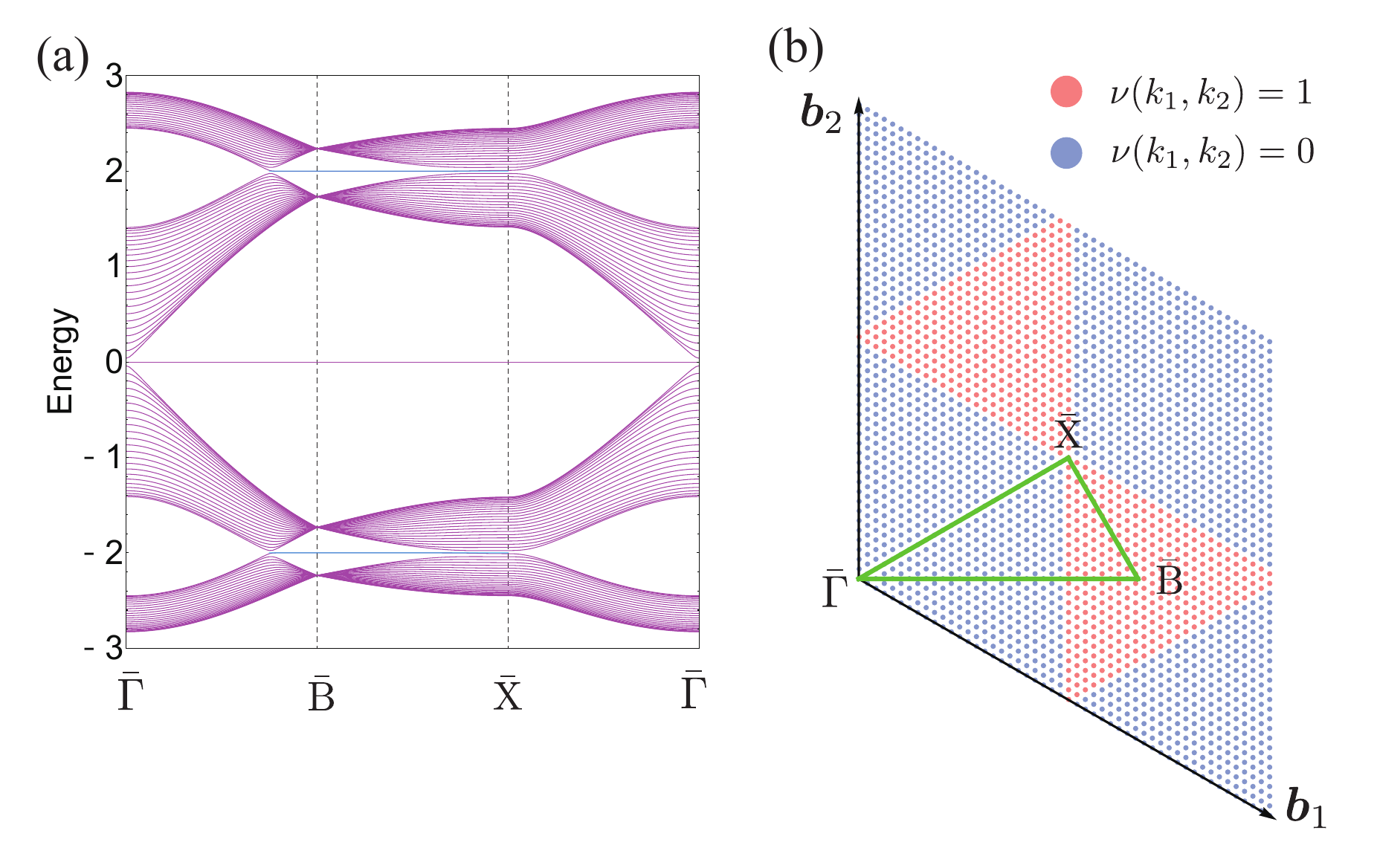}
\caption{(a) The dispersions of the decorated diamond model 
with a slab geometry. The cyan lines represent the flat surface modes.
(b) The map of the winding number defined for $q^{\rm (D)}_{\bm{k}}$ in Eq.~(\ref{eq:winding_SD}).
We set $t = 1$. }
  \label{fig:sd_surf}
 \end{center}
\end{figure}

\subsection{Model and bulk properties}
Next, we consider the decorated diamond model~\cite{Attig2017} [Fig.~\ref{fig:sd_bulk}(a)],which is a composite lattice of the diamond and the pyrochlore lattices,
as an example of 
the SR nodal-line semimetal. 
In this model, 
the number of sublattices is six; two of them come from the diamond lattice and the rest four from the pyrochlore lattice.
Hence, $\Psi_{\bm{k}}$, introduced in Sec.~\ref{sec:gen},
becomes $4\times 2$ matrix given as
\begin{eqnarray}
\Psi_{\bm{k}} = 
\begin{pmatrix}
1 & 1 \\
e^{i\bm{k}\cdot \bm{a}_1} & 1 \\
e^{i\bm{k}\cdot \bm{a}_2} & 1 \\
e^{i\bm{k}\cdot \bm{a}_3} & 1 \\
\end{pmatrix},
\end{eqnarray}
where $\bm{a}_1 = \left(0, \frac{1}{2},\frac{1}{2} \right)$, $\bm{a}_2 = \left( \frac{1}{2},0, \frac{1}{2} \right)$,
$\bm{a}_3 = \left( \frac{1}{2}, \frac{1}{2},0 \right)$ are lattice vectors;
the corresponding reciprocal lattice vectors are
$\bm{b}_1 = 2 \pi \left(-1,1,1 \right)$, $\bm{b}_2 = 2\pi \left(1,-1,1 \right)$,
$\bm{b}_3 = 2 \pi  \left(1,1,-1\right)$.

The square of $\mathscr{H}_{\bm{k}}$ is 
a direct sum of the diamond model and the pyrochlore model:
\begin{eqnarray}
\mathscr{H}_{\bm{k}}^2  = 
\begin{pmatrix}
\mathscr{H}_{\bm{k}}^{(\mathrm{D})} &\mathcal{O}_{2,4}  \\
\mathcal{O}_{4,2} & \mathscr{H}_{\bm{k}}^{(\mathrm{P})}  \\
\end{pmatrix}.
\end{eqnarray}
The matrix $\mathscr{H}_{\bm{k}}^{(\mathrm{D})}$
corresponds to the diamond lattice model with the NN hopping $t^2$ and the on-site potential $4t^2$;
similarly, $\mathscr{H}_{\bm{k}}^{(\mathrm{P})}$
corresponds to the diamond lattice model with the NN hopping $t^2$ and the on-site potential $2t^2$.
We again define the quantity corresponding to the $(2,1)$ component of $\mathscr{H}_{\bm{k}}^{(\mathrm{D})}$:
\begin{eqnarray}
q^{\rm (D)}_{\bm{k}} = 1 + \sum_{j=1}^3 e^{i \bm{k} \cdot \bm{a}_j},
\end{eqnarray}
which we will use in the next subsection.

In Fig.~\ref{fig:sd_bulk}(b), we plot the bulk band structure.
We see nodal lines between the X and W points, which are inherited from the diamond and pyrochlore models. 

\subsection{Surface states and topological protection}
Let us investigate the boundary modes of this model. 
Consider a slab geometry whose surface is parallel to the plane spanned by $\bm{a}_1$ and $\bm{a}_2$. 
In this case, 
$k_1$ and $k_2$ remain as good quantum numbers, while $k_3$ is not. 
In Fig.~\ref{fig:sd_surf}(a), we plot the band structure for the slab on the high-symmetry lines of the surface Brillouin zone
[green lines in Fig.~\ref{fig:sd_surf}(b)].
We see that the flat surface states appear, whose energies are $\pm 2|t|$. 

As discussed in Sec.~\ref{sec:gen}, the topological protection of the surface state can be dictated by 
calculating the following winding number as a function of $k_1$ and $k_2$: 
\begin{eqnarray}
\nu (k_1,k_2) =\frac{1}{2\pi i} \int_0^1 d k_3 \hspace{0.5mm} \frac{\partial_{k_3} q^{\rm (D)}_{\bm{k}}}{q^{\rm (D)}_{\bm{k}} }. \label{eq:winding_SD}
\end{eqnarray}
Performing the integration in the same manner as in Eqs.~(\ref{eq:winding_H}) and (\ref{eq:winding_H_2}), 
we find that $\nu (k_1,k_2)$ takes 0 ($1$) if $|1 + e ^{i 2\pi k_1}  + e ^{i 2\pi k_2}  |> 1$ 
($|1 + e ^{i 2\pi k_1}  + e ^{i 2\pi k_2} |< 1$). 
For the critical case where $|1 + e ^{i 2\pi k_1}  + e ^{i 2\pi k_2} | =  1$, 
i.e., $k_1 = \frac{1}{2}$, $k_2 = \frac{1}{2}$, and $k_1 - k_2 = \pm \frac{1}{2}$,
the band gap closes at some $k_3$, which correspond to the nodal lines. 
Similarly to the decorated honeycomb lattice, 
this result indicates the topological protection of the bulk nodal lines and the surface states.
Namely, 
the finite winding number indicates the existence of zero-energy surface state for $\mathscr{H}_{\bm{k}}^{(\mathrm{D})}- 4t^2$,
hence the the surface states whose energies are $\pm 2 |t|$ appear for the decorated diamond model.
Figure~\ref{fig:sd_surf}(b) shows the map of the winding number as a function of the surface wavevectors. 
Clearly, the above correspondence between the surface states and the winding number can be confirmed. 

\section{Realization in a spring-mass model \label{sec:SMmodel}}
In this section, we argue the realization of the SR-TSM we have discussed so far in a spring-mass model.
The aims of studying the spring-mass model are (i) to propose an experimentally feasible setup for the SR-TSMs,
and (ii) to demonstrate the topological stability of the SR-TSM against the symmetry-preserving perturbations. 
Concerning (ii), we note that the symmetry-preserving perturbations can be achieved by changing
the strength of the coupling between longitudinal and transverse modes, which is inherent in spring-mass models.

\begin{figure}[b]
\begin{center}
\includegraphics[clip,width = 0.98\linewidth]{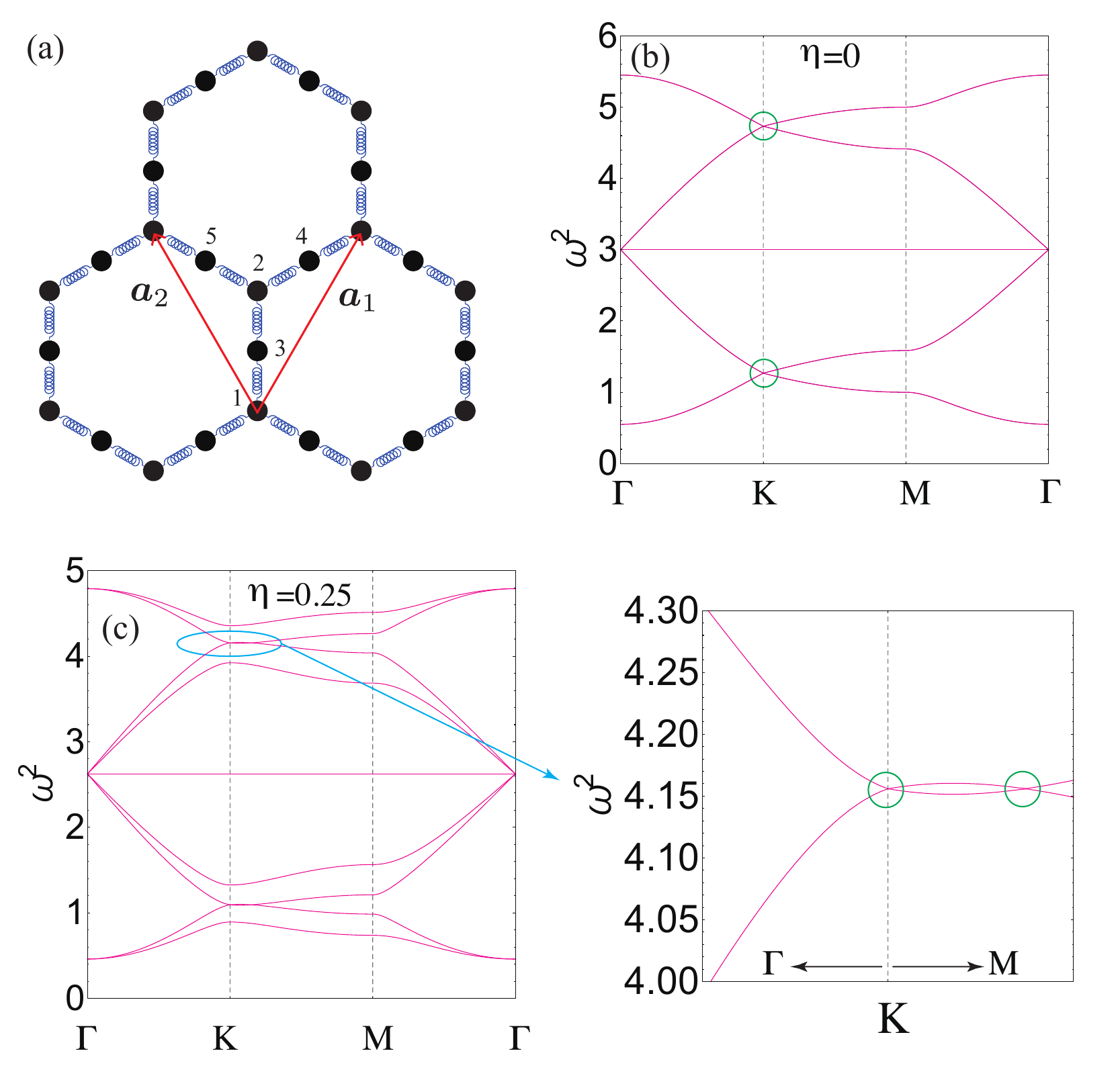}
\caption{
(a) A schematic figure of the spring-mass decorated honeycomb model.
We have introduced potential force arising from the dents of the floor 
so that the diagonal elements of $\Gamma(\bm{k})$ become $[\Gamma(\bm{k})]_{ii}=3\kappa(1-\eta/2)$ for $i=1,\cdots,10$.
Dispersion relations of the spring-mass decorated honeycomb model
for $\kappa=m=1$ with (b) $\eta = 0$ and (c) $\eta = 0.25$.
Green circles denote the finite-energy Dirac points. }
\label{fig:SMmodel}
\end{center}
\end{figure}

We consider a system of mass points and springs aligned on the decorated honeycomb lattice
[Fig.~\ref{fig:SMmodel}(a)].
Each mass point can move in any direction of two-dimensional space, in the vicinity of the stationary point.
The mass points on sublattices 3, 4, and 5 are placed on dents of the floor which cause the gravitational potential~\cite{Kariyado2016,Yoshida2019,Wakao2020_2}.
The Lagrangian of the system is given as
\begin{eqnarray}
\mathcal{L}&=& T-U_{\mathrm{g}}-U_{\mathrm{sp}}, \label{eq:Lag}
\end{eqnarray}
with 
\begin{eqnarray}
T &=& \frac{m}{2} \sum_{\bm{R}} \left(  \dot{x}_{\bm{R},\mu} \right)^2, \label{eq:kin}
\end{eqnarray}
\begin{eqnarray}
U_{\mathrm{g}} &=& \sum_{\bm{R}} g_{\bm{R},\mu\nu} x_{\bm{R},\mu}  x_{\bm{R},\nu}, \label{eq:Ug}
\end{eqnarray}
and 
\begin{eqnarray}
U_{\mathrm{sp}} &=& \frac{\kappa}{2} \sum_{\langle \bm{R}, \bm{R}' \rangle} (x_{\bm{R},\mu}-x_{\bm{R},\mu}) \gamma_{\bm{R}-\bm{R}',\mu\nu} (x_{\bm{R}',\nu}-x_{\bm{R}',\nu}). \nonumber \\  \label{eq:Usp}
\end{eqnarray}
Here, we have assumed the summation over repeated indices $\mu$ and $\nu$ ($\mu,\nu=x,y$),
and the dot in Eq.~(\ref{eq:kin}) stands for the time derivative.
In Eq.~(\ref{eq:Lag}), $T$ is the kinetic energy, 
$U_{\mathrm{g}}$ is the potential energy from the dents on the floor,
and $U_{\mathrm{sp}}$ is the potential energy describing the restoring force of the springs.
The vector $\bm{R}$ specifies the sites which forms a decorated honeycomb lattice.
The natural length of a spring is denoted by $l_0$. 
The ratio of between distance of neighboring sites (specified by $\bm{R}$ and $\bm{R}'$) 
and $l_0$ is denoted by $\eta=l_0/|\bm{R}-\bm{R}'|$, which determines the tension of the spring.
The distance between neighboring sites is chosen as the unit of length.
The vector $\bm{x}_{\bm{R}}=(\bm{x}_{\bm{R},x},\bm{x}_{\bm{R},y})$ describes the displacement of the mass point at the site $\bm{R}$.

As for the matrix $g_{\bm{R}}$ in Eq.~(\ref{eq:Ug}), we set
\begin{eqnarray}
g_{\bm{R}} =  \left\{
\begin{array}{cc}
0 & \mathrm{if} \hspace{0.5mm} {\bm{R} \in 1,2} \\
\kappa \left(1 - \frac{1}{2} \eta \right) \tau_0 +  \kappa \eta \tau_z & \mathrm{if} \hspace{0.5mm} {\bm{R} \in 3}\\
\kappa \left(1 - \frac{1}{2} \eta \right) \tau_0 -  \frac{\kappa \eta}{2}\left(\sqrt{3}\tau_x + \tau_z\right) & \mathrm{if} \hspace{0.5mm} {\bm{R} \in 4}\\
\kappa \left(1 - \frac{1}{2} \eta \right) \tau_0 - \frac{\kappa \eta}{2}\left(- \sqrt{3}\tau_x + \tau_z\right) & \mathrm{if} \hspace{0.5mm} {\bm{R} \in 5}\\
\end{array}
\right. , 
\nonumber \\
\end{eqnarray}  
with $\tau_0$ being the $2\times 2$ identity matrix and $\bm{\tau} = \left(\tau_x, \tau_y,\tau_z \right)$ being Pauli matrices. 
Note that the potential is introduced so that 
the eigenvalue problem is equivalent to that for 
the tight-binding model in the strong tension limit~\cite{Kariyado2016,Yoshida2019,Wakao2020_2}.
Further, $g_{\bm{R}}$ is also essential for restoring the chiral symmetry for the squared systems, which is essential for realizing the SR-TSM in this system. This implies that, for three-dimensional systems where the systems are not placed on the floor,
the realization of the SR-TSM is not straightforward.
We also note that $\eta$ has to be smaller than $\frac{2}{3}$ 
so that the dents have an ellipsoidal surface (i.e., all the eigenvalues of $g_{\bm{R}}$ are positive).

The matrix $\gamma$ in (\ref{eq:Usp}) is a $2\times 2$ matrix whose elements is defined as
\begin{eqnarray}
\gamma_{\delta \bm{R},\mu\nu}&=& (1-\eta)\delta_{\mu\nu} + \eta \widehat{\delta \bm{R}}_\mu \widehat{\delta \bm{R}}_\nu,\label{eq:gamma_real}
\end{eqnarray}
with $\widehat{\delta \bm{R}}=\delta \bm{R}/|\delta \bm{R}|$ and $\delta \bm{R}$ being the vector connecting the neighboring sites, respectively.

The normal modes of the frequency $\omega$ are obtained as follows.
Writing $\bm{R} = \bar{\bm{r}} + \bm{r}_{\alpha}$
where $\bar{\bm{r}}$ and $\bm{r}_{\alpha}$ are the position of the unit cell and the position of the sublattice $\alpha$ of which $\bm{R}$ belongs to, respectively, 
we apply the Fourier transformation:
\begin{eqnarray}
x_{\bm{R},\mu} = \frac{1}{N} \sum_{\bm{k}} u^{\alpha}_{\bm{k},\mu } e^{i \bm{k} \cdot \bar{\bm{r}}}.
\end{eqnarray}
Then, from the Euler-Lagrange equation 
\begin{eqnarray}
\frac{d}{dt} \left( \frac{\delta \mathcal{L}}{\delta \dot{u}^{\alpha}_{-\bm{k},\mu}} \right) - \frac{\delta \mathcal{L}}{\delta u^{\alpha}_{- \bm{k},\mu}  } =0, 
\end{eqnarray}
we obtain the equation of motion in the momentum space:
\begin{eqnarray}
\ddot{\bm{u}}_{\bm{k}}&=& -\Gamma(\bm{k}) \bm{u}_{\bm{k}},
\end{eqnarray}
with
$\bm{u}_{\bm{k}} 
= \left(u^1_{\bm{k},x},u^1_{\bm{k},y}, \cdots ,u^5_{\bm{k},x},u^5_{\bm{k},y} \right)^{\rm T}$
and
\begin{eqnarray}
\Gamma (\bm{k}) 
&=& 3\frac{\kappa}{m} \left(1 -\frac{\eta}{2} \right) I_{10}  \nonumber \\
&-& \frac{\kappa}{m}
\begin{pmatrix}
0 &0 & \gamma^{(1,3)}_{\bm{k}} &   \gamma^{(1,4)}_{\bm{k}},  \gamma^{(1,5)}_{\bm{k}} \\
0 & 0 & \gamma^{(2,3)}_{\bm{k}} &   \gamma^{(2,4)}_{\bm{k}},  \gamma^{(2,5)}_{\bm{k}} \\
\gamma^{(1,3),\dagger}_{\bm{k}}  &  \gamma^{(2,3),\dagger}_{\bm{k}}& 0 & 0 & 0 \\
\gamma^{(1,4),\dagger}_{\bm{k}}  &  \gamma^{(2,4),\dagger}_{\bm{k}}& 0 & 0 & 0 \\
\gamma^{(1,5),\dagger}_{\bm{k}}  &  \gamma^{(2,5),\dagger}_{\bm{k}}& 0 & 0 & 0 \\
\end{pmatrix}.\nonumber \\
\label{eq:gamma_mat_k}
\end{eqnarray}
Here, $\gamma^{(\alpha,\beta)}_{\bm{k}}$ are $2 \times 2$ matrices obtained by performing the Fourier transformation
of $\gamma$ in Eq.~(\ref{eq:gamma_real}).
Their explicit forms are given in Appendix~\ref{sec:gamma}.
The matrix $\Gamma (\bm{k})$ is referred to as the dynamical matrix in the literature. 
One can show that the square of the second term of Eq.~(\ref{eq:gamma_mat_k}) preserves the chiral symmetry in the honeycomb subspace,
which is necessary for realizing the SR-TSM.
By further assuming the relation 
$\bm{u}_{\bm{k}}(t)=e^{i\omega t} \bm{\phi}_{\bm{k}}$, we have 
\begin{eqnarray}
\omega^2 \bm{\phi}_{\bm{k}} = \Gamma (\bm{k}) \bm{\phi}_{\bm{k}}.
\end{eqnarray}
Solving the above eigenvalue equation, we have the dispersion relation for the spring-mass system.

Figure~\ref{fig:SMmodel}(b) shows the dispersion relation for $\eta=0$.
Note that, in this case, all of the matrices $\gamma^{(\alpha,\beta)}_{\bm{k}}$ are diagonal,
and $\Gamma (\bm{k})$ is equivalent to two copies of the tight-binding Hamiltonian (up to the constant shift).
In this figure, we can see that the band structure 
of the tight-binding model is reproduced up to the constant shift $3\kappa/m$ which does not matter for the topological properties.
We note that for $\eta=0$ the transverse and longitudinal modes are decoupled and each band is doubly degenerate.
Namely, each of the Dirac cones for $\omega^2\sim 4.7$ and $\omega^2\sim 1.2$ at the K point is doubly degenerate, 
which is consistent with the fact that the winding number of $\Gamma(\bm{k})-3\frac{\kappa}{m} I_2$ is two.

For a finite value of $\eta$, the transverse and longitudinal modes are coupled. 
As a result, some of the matrices $\gamma^{(\alpha,\beta)}_{\bm{k}}$ become off-diagonal, and 
$\Gamma (\bm{k})$ is deviated from two copies of the tight-binding Hamiltonian.
In this case, we find that the doubly degenerate Dirac cone with the winding number two 
is not gapped out but splits into two Dirac cones each of which has the winding number one,
as shown in Fig.~\ref{fig:SMmodel}(c). 
More precisely, one of the Dirac cones remains at the K point, while the other is placed on 
the K-M line. 
This behavior, which is reminiscent of the bilayer graphene with AA stacking~\cite{Andres2008,Rakhmanov2012}, is also seen in the spring-mass model on a conventional honeycomb lattice~\cite{Kariyado2015}.
This result indicates that the Dirac cones are not gapped out by the inclusion of the intermode coupling,
which manifests the topological stability of the SR-TSM against the symmetry-preserving change of parameters.

\section{Summary\label{sec:summary}}
We have proposed the SR-TSM,
where topologically-protected point or line nodes are inherited from the squared Hamiltonian. 
As concrete examples, 
we study the decorated honeycomb model and the decorated diamond model,
where the SR Dirac semimetal and SR nodal-line semimetal are realized, respectively. 
There, the Dirac cones and nodal lines appear at finite energy, 
and they are protected by the winding number defined for the squared Hamiltonian.
We can also see the bulk-boundary correspondence between this winding number and the finite-energy flat edge or surface modes. 

We have further proposed that the SR-TSM can be realized in a spring-mass model 
with the decorated honeycomb arrangement. 
We have found that the finite-energy band touching points 
are robust against the change of the tension parameter $\eta$,
which indicates their topological protection by the winding number of the parental honeycomb model.
We expect that the finite-energy gapless points in the bulk and the flat edge modes associated with them
will be observed experimentally~\cite{Huber2016}, by implementing the decorated honeycomb structure.

Before closing, we make two remarks which are related to topological classification.
First, for the generalization of the dimensionality, the SR-TSMs in class AIII 
are obtained when the codimension of the nodes~\cite{Remark_codim} is
even because the conventional TSMs in the same class are found under the same condition~\cite{Chiu2014}. 
Second, extension of SR-TSMs to generic topological classes will be straightforward, 
namely they are obtained by performing the square-root operation to other classes of TSMs.
In such systems, the finite-energy gapless nodes are protected by the symmetries of the squared Hamiltonian.
In fact, finite-energy gapless points and lines appear in various condensed-matter systems~\cite{Alidoust2018}.
As well as explicit classification results, the detailed analysis of these edge states is left as a future work to be addressed.
We hope that the SR-TSM proposed here provides a renewed perspective on topological band strictures.

\acknowledgments
We wish to thank Yoshihito Kuno for bringing our attention to the square-root topological insulators,
and for the collaboration in the earlier work (Ref.~\cite{Mizoguchi2020_sq}).
This work is supported by the JSPS KAKENHI, 
Grant Numbers JP17H06138, JP20K14371 (T. M.), JP20H04627 (T. Y.), MEXT, Japan.

\appendix
\section{Stability against on-site potential \label{sec:onsite}}
\begin{figure}[b]
\begin{center}
\includegraphics[clip,width = 0.98\linewidth]{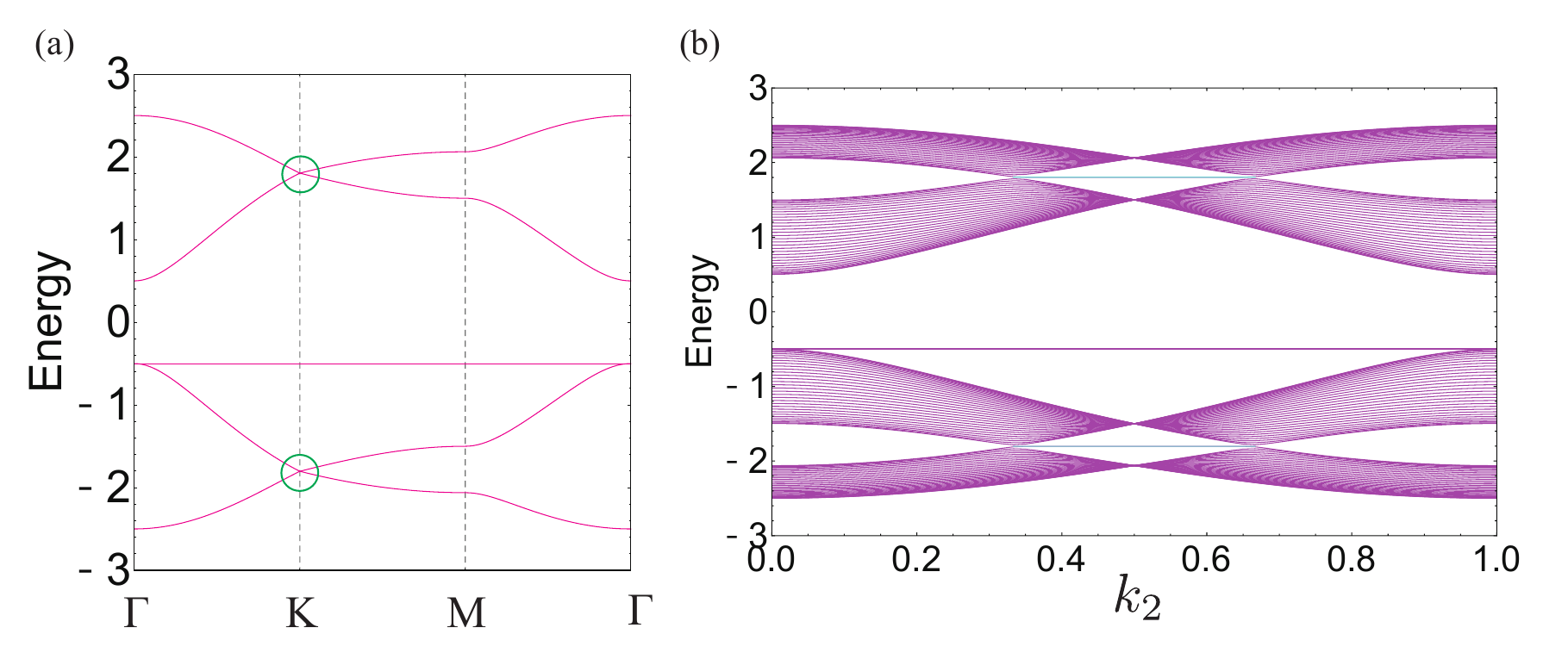}
\caption{(a) The band structure for the decorated honeycomb model with $t = 1$ and $V=0.5$.
The finite-energy Dirac points are denoted by green circles.
(b) The dispersions for the cylinder with the same parameters.
The cyan lines represent the flat edge modes. }
  \label{fig:model_dh_onsite}
 \end{center}
\end{figure}
We show that the SR-TSM is stable against the on-site potential proportional to the chiral operator $\Gamma$ of Eq.~(\ref{eq:Gamma}).
Consider the Hamiltonian 
\begin{eqnarray}
\mathscr{H}^{\prime}_{\bm{k}} =
\begin{pmatrix}
V I_{N} & \Psi_{\bm{k}}^\dagger \\
\Psi_{\bm{k}}& -VI_{M} \\
\end{pmatrix},
\end{eqnarray}
where $V$ is the strength of the on-site potential.
Importantly, the square of $\mathscr{H}^{\prime}_{\bm{k}}$ is still block-diagonalized as
\begin{eqnarray}
\left(\mathscr{H}^{\prime}_{\bm{k}}\right)^2 =
\begin{pmatrix}
V^2 I_{N} + \Psi_{\bm{k}}^\dagger \Psi_{\bm{k}}& \mathcal{O}_{N,M} \\
\mathcal{O}_{M,N}  &V^2 I_{M} + \Psi_{\bm{k}} \Psi^\dagger_{\bm{k}} \\
\end{pmatrix}.
\end{eqnarray}
This indicates that the on-site potential proportional to the chiral operator 
causes a mere constant shift to the squared Hamiltonian.
Consequently, the topological nature of the Hamiltonian without on-site potential is unchanged. 

To confirm this, we calculate the band structures of bulk and 
cylinder in the decorated honeycomb model with on-site potential.
The results are shown in Fig.~\ref{fig:model_dh_onsite},
where we clearly see that the finite-energy Dirac cones in bulk and the flat edge modes in cylinder,
indicating that the stability of the SR-TSM against the on-site potential. 

\section{Details of the dynamical matrix \label{sec:gamma}}
Here we list the forms of $\gamma^{(\alpha,\alpha^\prime)}_{\bm{k}}$ in Eq.~(\ref{eq:gamma_mat_k}):
\begin{subequations}
\begin{eqnarray}
\gamma^{(1,3)}_{\bm{k}}= 
\begin{pmatrix}
1-\eta & 0\\
0 & 1 \\
\end{pmatrix},
\end{eqnarray}
\begin{eqnarray}
\gamma^{(1,4)}_{\bm{k}} = e^{-i\bm{k} \cdot \bm{a}_1}
\begin{pmatrix}
1-\frac{\eta}{4} & \frac{\sqrt{3}\eta}{4}\\
 \frac{\sqrt{3}\eta}{4} & 1-\frac{3 \eta}{4} \\
\end{pmatrix},
\end{eqnarray}
\begin{eqnarray}
\gamma^{(1,5)}_{\bm{k}} = e^{-i\bm{k} \cdot \bm{a}_2}
\begin{pmatrix}
1-\frac{\eta}{4} & - \frac{\sqrt{3}\eta}{4}\\
 - \frac{\sqrt{3}\eta}{4} & 1-\frac{3 \eta}{4} \\
\end{pmatrix},
\end{eqnarray}
\begin{eqnarray}
\gamma^{(2,3)}_{\bm{k}} = 
\begin{pmatrix}
1-\eta & 0\\
0 & 1 \\
\end{pmatrix},
\end{eqnarray}
\begin{eqnarray}
\gamma^{(2,4)}_{\bm{k}} = 
\begin{pmatrix}
1-\frac{\eta}{4} & \frac{\sqrt{3}\eta}{4}\\
 \frac{\sqrt{3}\eta}{4} & 1-\frac{3 \eta}{4} \\
\end{pmatrix},
\end{eqnarray}
and 
\begin{eqnarray}
\gamma^{(2,5)}_{\bm{k}} = 
\begin{pmatrix}
1-\frac{\eta}{4} & - \frac{\sqrt{3}\eta}{4}\\
 - \frac{\sqrt{3}\eta}{4} & 1-\frac{3 \eta}{4} \\
\end{pmatrix}.
\end{eqnarray}
\end{subequations}

\bibliographystyle{apsrev4-2}
\bibliography{SQroot_TSM}

\begin{thebibliography}{63}
\bibitem [{Rem({\natexlab{a}})}]{Remark}%
  \BibitemOpen
  \href@noop {} {}\bibinfo {note} {Note that K and K$^\prime$ points correspond
  to $(k_1,k_2) = \left( \frac{1}{3}, \frac{2}{3}\right)$ and $\left(
  \frac{2}{3}, \frac{1}{3}\right)$, respectively.}\BibitemShut {Stop}%
\bibitem [{Rem({\natexlab{b}})}]{Remark_codim}%
  \BibitemOpen
  \href@noop {} {}\bibinfo {note} {The co-dimension is defined as $D-N$, where
  $D$ is the dimensions of the bulk and $N$ is that of the node.}\BibitemShut
  {Stop}%
\end{thebibliography}

\end{document}